# Crystal growth, structure and ferromagnetic properties of a $Ce_3Pt_{23}Si_{11}$ single crystal

C. Opagiste*, C. Paulsen, E. Lhotel, P. Rodière, R.-M. Galera, P. Bordet and P. Lejay

Institut Néel, CNRS et Université Joseph Fourier, BP166, F-38042 Grenoble Cedex 9, France

**Abstract**

A high-quality single crystal of $Ce_3Pt_{23}Si_{11}$ has been grown using the Czochralski method. The crystal structure is presented and the chemical composition has been checked using an electron microprobe analyzer. Measurements of the electrical resistivity and magnetic susceptibility performed at low temperature show a ferromagnetic transition at Tc = 0.44 K.

*****Corresponding author**:        Christine Opagiste

e-mail: christine.opagiste@grenoble.cnrs.fr

**I. Introduction**

The discovery of the non centro-symmetric heavy fermion superconductor $CePt_3Si$ [1] has launched both theoretical and experimental research on a new type of unconventional superconductor [2,3]. Moreover, while antiferromagnetism and superconductivity coexistence in $CePt_3Si$ at low temperature, both transitions are sample dependent. Several reports on the magnetic properties, superconductivity and specific heat measurements of $CePt_3Si$ have shown that certain heat treatments, or slight off-stoichiometries on the Ce, Pt and Si sites, have a great influence on the physical properties of the $CePt_3Si$ phase [4,5]. It is therefore interesting to study carefully the ternary Ce-Pt-Si system.

An isothermal section of the phase diagram at 600°C of this ternary system have been presented by Gribanov et al. [6]. This study confirmed the existence of a new compound, namely $Ce_3Pt_{23}Si_{11}$, which could be found in presence of $CePt_3Si$ when the samples are slightly Ce deficient. The chemical composition of this compound had also been previously reported to correspond to the formula $Ce_2Pt_{15}Si_7$ [7]. This compound is isotypic to its U counterpart in the U-Pt-Si ternary system [8-10]. The presence of $Ce_3Pt_{23}Si_{11}$ was pointed out by several physical studies performed on $CePt_3Si$, and was suspected to interfere with superconducting properties of $CePt_3Si$ [7,11]. One recent study made by Kundaliya et al. [12] on $Ce_3Pt_{23}Si_{11}$ polycrystalline samples shows a Curie-Weiss behavior with $\mu_{eff} = 2.48\mu_B$ and $\theta_p = -4$ K. The specific heat shows a large upturn below 0.6 K indicative of magnetic ordering in this compound. Kim et al. [7] have performed specific heat measurements on a polycrystalline sample and they presumed that $Ce_3Pt_{23}Si_{11}$ underwent an antiferromagnetic transition at 0.41 K. Up to now, no physical investigation has been carried out on a pure $Ce_3Pt_{23}Si_{11}$ single crystal at very low temperatures. In this paper we describe physical properties of a very high quality single crystal of $Ce_3Pt_{23}Si_{11}$ down to 100 mK.

## II. Crystal growth

In the first step of the sample preparation, a polycrystalline sample of $Ce_3Pt_{23}Si_{11}$ was prepared in an induction furnace, in a cold copper crucible and in a high purity argon atmosphere, with a stoichiometric amount of Ce (99.99%, Johnson Matthey), Pt (99.95%, Alfa Aesar) and Si (99.9999%, Alfa Aesar). The sample was melted several times to improve homogeneity. Mass losses during this first step were less than 0,1 %. The sample was checked by conventional X-ray powder diffraction and appeared to be single phase (fig. 1).

The polycrystalline sample was then placed in a three arc furnace equipped with a Czochralski puller, and again in a purified argon atmosphere. A polycrystalline pulling tip of $Ce_3Pt_{23}Si_{11}$ was used to start crystal growth. The growth parameters were the followings: the translation speed was 20 mm/h and the rotation speed was 20 rpm. After one or two necking operations, the tip became a single crystal seed. The final crystal was cylindrical shaped, with a diameter of about 1.5 mm and a length of 10 mm as shown in figure 2a.

## III. Results and discussion

### III-1 Crystallographic analysis

The crystal was checked and oriented using the back-scattering X-ray Laue technique performed on a PANanalytical PW3830 diffractometer. The growth direction was found to be parallel to the [111] crystallographic direction.

Scanning Electron Microscope (SEM) analyses were made using a JEOL 840A microscope. The results confirmed that the crystal is single phase and that it has a very good stoichiometric homogeneity. Figure 2a shows a typical single crystalline ingot of $Ce_3Pt_{23}Si_{11}$ obtained with the Czochralski technique. Figure 2b shows homogenous grey electron back scattering obtained by SEM. Because of the inherent difficulties to determining with a high accuracy both heavy and light elements, we performed microprobe analysis using a $CePt_2Si_2$ single crystal [13] as standard. These microprobe analyses were performed using a Cameca SX50 electron microprobe analyser. Our results (see table 1) and the fact that we have a small standard deviation, indicate that the crystal formula was still ambiguous after these tests.

Therefore, in order to confirm the crystal structure and check the chemical composition of the samples, a single crystal was studied by x-ray diffraction, using a KappaApexII Bruker diffractometer equipped with graphite monochromatized $AgK\alpha$ radiation. The sample was cut from a small crystal piece, with an approximately triangular shape of maximum length 0.16mm. The experimental data are summarized in Table 2. The data collection was carried out at room temperature up to $sin\theta/\lambda=1$, with a redundancy of about 18, yielding 20392 reflections out of which 12100 were considered as observed (i.e. $I>3\sigma(I)$; $\sigma(I)$ corresponds to the estimated standard deviations of individual intensities I known from the measurement). The data reduction was carried out using the maXus package [14] and refinements with the Jana2000 program [15]. Anisotropic absorption correction was carried out using sadabs. The corrected data were average in space group Fm-3m leading to 1138 unique reflections of which 897 were observed, with Rint=8.5%. Such a high value is not uncommon in heavily absorbing samples ($\mu=75.07mm^{-1}$), especially when having a non crystallographic or spherical shape, as it is the case here. The cell parameter value was obtained by post-refining the positions of 2774 reflections. It is noticeably larger than the one reported by Tursina et al. (16.901 Å compared to 16.837 Å) [16]. Since this difference does not result from compositional variation as seen by the closeness of the two refinement results, it may most likely be attributed to experimental inaccuracy due to the heavy absorption. In this respect, the value determined by the Rietveld refinement of the powder diffraction data is probably less sensitive to such effects and therefore more accurate.

The structure was readily solved by direct methods and the refinements including anisotropic thermal parameters for all atoms lead to results very close to those of Tursina et al. [16] (see Table 3, 4). An isotropic extinction correction was included in the refinement (type I, lorentzian distribution). In order to account for the difference between the starting composition and the crystallographic stoichiometry, we refined the occupancy of the cerium site. As seen on the table, it does not deviate from unity by a sizeable amount compared to the experimental precision, and its refinement did not lead to any improvement of the agreement factors. Based on all of these considerations, we conclude that the sample is most likely stoichiometric and its chemical composition is given by the crystallographic multiplicities of the occupied sites, i.e. $Ce_3Pt_{23}Si_{11}$.

### III-2 Electrical resistivity measurement

A long bar of section $S=0.1mm^2$ and length $l=1mm$ along the [100] axis was cut from the large single crystal and then annealed at 900°C for 7 days under an ultra high vacuum furnace in order to relax internal stress.

The resistivity measurements were made using the conventional four probe technique and a low impedance commercial bridge (TRMC2 with F-card) operating at 25Hz. Gold wires (10μm diameter) were spark-welded directly onto the sample so as to minimize the contact resistance. The electrical resistivity was measured along the [100] crystallographic directions in the temperature range 70 mK – 300 K. At the lowest temperature, the absence of self heating was checked for the low current excitation of 1mA used in the measurements.

At room temperature the resistivity is 83.2 μΩcm. A second sample with a less favourable form factor gave a similar room temperature resistivity, within the experimental error bars (5%). Fig. 3 shows the temperature dependence of the resistivity from 10K down to 70mK. A sharp kink is observed at 0.45K with a strong decrease of the resistivity. Below 200 mK the resistivity saturates to a value of 12.16 μΩ.cm. This gives a residual resistivity ratio (RRR) of 6.8. The kink suggests the presence of an ordered state below 0.45K, decreasing the scattering rate at the proximity of the phase transition. In order to get more information about the transition, we have performed magnetic measurements.

### III-3 Magnetic measurements

A single crystal of $Ce_3Pt_{23}Si_{11}$ was cut from the middle of the melt and was annealed in the same conditions as mentioned before. The sample had the shape of a truncated cone: the length was approximately 4.86 mm, and the diameter at the top of the cone was 0.93 mm, and at the bottom 1.32 mm. The mass was 61.2 mg

Figure 4 shows results of magnetisation measurements from room temperature down to 1.9 K using a commercial Quantum Design MPMS SQUID magnetometer. The measurements were made along two different crystallographic directions ( [100] and [111] ). The measurements for the two directions show no significant difference as expected for a cubic system in the paramagnetic phase. The magnetic susceptibility of figure 4 is best describe as a Curie Weiss paramagnetic behaviour: $\chi(T) = \frac{C}{T - \theta_p}$. A fit of the susceptibility data over the whole temperature range yields an effective moment per Ce atom of $\mu_{eff}$ = 2.53 $\mu_B$. This value corresponds to the theoretical free $Ce^{3+}$ ion ($\mu_{eff}$ = 2.54 $\mu_B$). In addition, we found $\theta_p$ = -3K . These results are very close to those reported by Kundaliya et al. [12].

To reach a lower temperature range, the susceptibility and magnetization measurements were made using two low temperature SQUID magnetometers developed at the MCBT, Institute Néel. One of the magnetometers is equipped with an 8 Tesla superconducting magnet, whereas the other is designed for low field measurements (< 0.2 Tesla). Both use miniature dilution refrigerators capable of cooling samples down to 70 mK. Absolute values of the magnetization are made by the extraction method.

Two sets of measurements were made: one set using the high field magnetometer, was made with the applied field perpendicular to the long axis of the crystal, and the other using the low field magnetometer, was made with the long axis of the crystal parallel to the applied field. The later correspond to the field parallel to the [111] crystallographic direction,.

Figure 5 shows the AC and the DC susceptibility vs. temperature with the field perpendicular to the long axis. The AC field strength was 0.5 Oe rms at 18 Hz, and for the DC susceptibility, $\chi = M / H$, the applied field was 10 Oe. As the temperature is decreased, the AC susceptibility increases, and at a temperature of 0.44 K, a very sharp peak is observed in the real part of the ac susceptibility ($\chi'$), accompanied by an abrupt onset of a signal in the imaginary part of the susceptibility ($\chi''$). The DC susceptibility is also seen to increase, to the same maximum, but below 0.44 K, the dc susceptibility remains constant. The sharpness of the peak is indicative of a phase transition at the critical temperature Tc= 0.44 K.

Figure 6 shows the AC susceptibility measured along the [111] crystallographic direction taken in an applied field of 0.1 Oe at 2 Hz. The peak at Tc=0.44 K is even more pronounced when measured along this direction, and in this small field. The imaginary part of the susceptibility shows the same abrupt onset at Tc, and a broad peak at lower temperature.

The exceptionally large magnitudes of the susceptibility at the peak for these two data sets are strong evidence of a ferromagnetic phase transition. The reason is as follows: it is well known that at a ferromagnetic transition, the intrinsic susceptibility, defined as $\chi_i = M / H_{int}$ diverges, where $H_{int} = H_a - NM$ is the internal field, $H_a$ is the applied field and N is the coefficient of demagnetization. In contrast, the applied susceptibility, $\chi_a = M / H_a$, as shown in Figure 5 and Figure 6, does not diverge, but reaches a maximum value which is simply $1/N$. The coefficient of demagnetization is a function of the shape of the sample, and depends on the axis being measured. Recall in cgs units, the sum of N along the principle axis of revolution of an ellipsoid is $4\pi$. For an ellipsoid of revolution and some other simple shapes N can be calculated exactly. However, in the present case, the sample was not an ellipsoid of revolution, but irregularly shaped, perhaps closer to a cylinder, so that at best we can only approximately assign values for N for the principal axis of the sample. Thus, for measurements perpendicular to the long axis, we estimate the coefficient using tables for a cylinder, to be $N_\perp = 5.7$. Consequently, the magnitude of the susceptibility would be $1/N_\perp = 0.175$ (emu/cm$^3$). Our data from Figure 5 show a value of $\chi_{peak} = 0.19$ emu/cm$^3$, thus slightly larger than our estimate. Parallel to the long axis we estimate $N_{//} = 1.1$, thus $1/N_{//} = 0.9$, and our measured value along this direction $\chi_{peak} = 1.1$ emu/cm$^3$ is again slightly larger than expected. Thus, considering the incertitude in estimating N for this irregular shaped sample, we can conclude that the sample undergoes a ferromagnetic phase transition.

A plot of the magnetization vs. field up to 8 Tesla is shown in Figure 7. These data were made with the field applied perpendicular to the long axis of the sample, at 3 different constant temperatures. At the lowest temperature shown, 90 mK, a rapid increase in the magnetization at very low field is followed by a gradual saturation of approximately 1.15 $\mu_B$ per cerium atom at 8 Tesla.

More evidence of a ferromagnetic phase transition can be seen in Figure 8 were plots of M vs. H (field parallel to [111] direction) for various fixed temperatures are shown, for temperatures above and below Tc. This is typical behaviour for a ferromagnetic material: above Tc, the m vs. H curves progressively become steeper as Tc is approached. At Tc, the initial slope of the magnetization vs. H reaches a maximum (in fact it is 1/N if the units are emu/cm$^3$). Below Tc, the initial slope is maintained, for larger applied fields, as the temperature decreases and each curve at lower and lower temperature, departs from the maximum slope at higher and higher fields. In fact, the departure from the maximum slope is an indication of the value of the spontaneous magnetization at that temperature.

Figure 9 shows a plot of the spontaneous magnetization as a function of temperature. The actual data points were taken from an Arrot plot [17] of the data of Figure 8. As can be seen, the spontaneous magnetization suddenly appears below Tc, and increases rapidly as the temperature is lowered. A fit of the data points just below Tc, to the power law scaling expression $M_{spontaneous} = M_{sat}(T-Tc)^{\beta}$ is also shown (solid line). The critical exponent from the fit was 0.38, and $M_{sat}=1.1$ $\mu_B$/Ce.

Fits of the data to critical scaling laws were made, but due to the unusual form of the sample, corrections for the demagnetization factors was problematic. In the future, a spherical sample will be prepared in order to study in more detail this particularly beautiful transition.

**IV. Conclusion**

We obtained high purity $Ce_3Pt_{23}Si_{11}$ crystal by the way of the Czochralski technique. Structural refinement performed on a single crystal confirmed that the sample is most likely stoichiometric. Electrical resistivity and magnetic properties were investigated from room temperature down to 100 mK. Resistivity and susceptibility measurements showed a ferromagnetic transition at 0.44 K. It is now clear that Ce deficient $CePt_3Si$ samples may contain $Ce_3Pt_{23}Si_{11}$ as impurity phase, which should dramatically affected both the magnetic and superconducting properties.

**Acknowledgments**

We greatly acknowledge the assistance of J. Balay in technical support, A. Hadj-Azzem in sample preparation and X-ray powder diffraction measurements and of S. Pairis in SEM examinations and microprobe analysis.

This work was supported by the Agence Nationale de la Recherche through the ICENET project : ANR-NT05-1_44475.

# Figure captions :

**Figure 1:** X-ray powder diffraction pattern (Cu-K$_\alpha$ radiation on a Philips PW1730 diffractometer) obtain for Ce$_3$Pt$_{23}$Si$_{11}$ polycrystalline sample. Circles mark the observed intensities and the solid line represents the best fit calculated from a Rietveld refinement. Vertical bars show ideal Bragg position. The bottom curve represents the difference between the observed and the calculated intensities.

**Figure 2:** Figure 2a shows a typical Ce$_3$Pt$_{23}$Si$_{11}$ crystal grown using the Czochralski technique. The crystal growth direction correspond to the [111] crystallographic axis. Figure 2b obtained with a scanning electron microscope demonstrates the perfect stoichiometric homogeneity of the crystal.

**Figure 3:** Electrical resistivity along the [100] axis direction versus temperature for Ce$_3$Pt$_{23}$Si$_{11}$ single crystal.

**Figure 4:** Magnetic susceptibility $\chi$ and the inverse susceptibility $\chi^{-1}$ (T) versus temperature from room temperature to 1.9 K. The solid line on the $\chi^{-1}$ (T) data is a fit based on the Curie-Weiss law.

**Figure 5:** Plot of the real ($\chi'$) and imaginary parts ($\chi''$) of the ac susceptibility, and the dc susceptibility ($\chi = M / H$) vs. temperature. The field was perpendicular to the long axis of the sample. The ac data were taken at 18 Hz using a field strength of 0.5 Oe rms. The applied field for the dc susceptibility was 10 Oe.

**Figure 6**: χ' and χ'' at 2 Hz, measured along the [111] crystallographic axis of the sample in a 0.1 Oe rms field. This sharpness of the peak, as well as exceptionally large magnitude of the susceptibility strongly imply that a ferromagnetic phase transition occurs at the critical temperature Tc= 0.44 K.

**Figure 7:** The magnetization plotted against field for fixed temperatures of 90 mK, 1 K and 4.2K. The sample seems to approach a saturation value of 1.15 $\mu_B$ per cerium atom.

**Figure 8** M vs. H (field parallel to [111] crystallographic axis) for various fixed temperatures above and below Tc

**Figure 9**. Plot of the spontaneous magnetization as a function of temperature. The points were taken from and Arrot plot of the data of Figure 8. The solid line is a fit to the scaling expression $M_{spontaneous} = (T-Tc)^{\beta}$.

Figure 1  (C. Opagiste et al.)

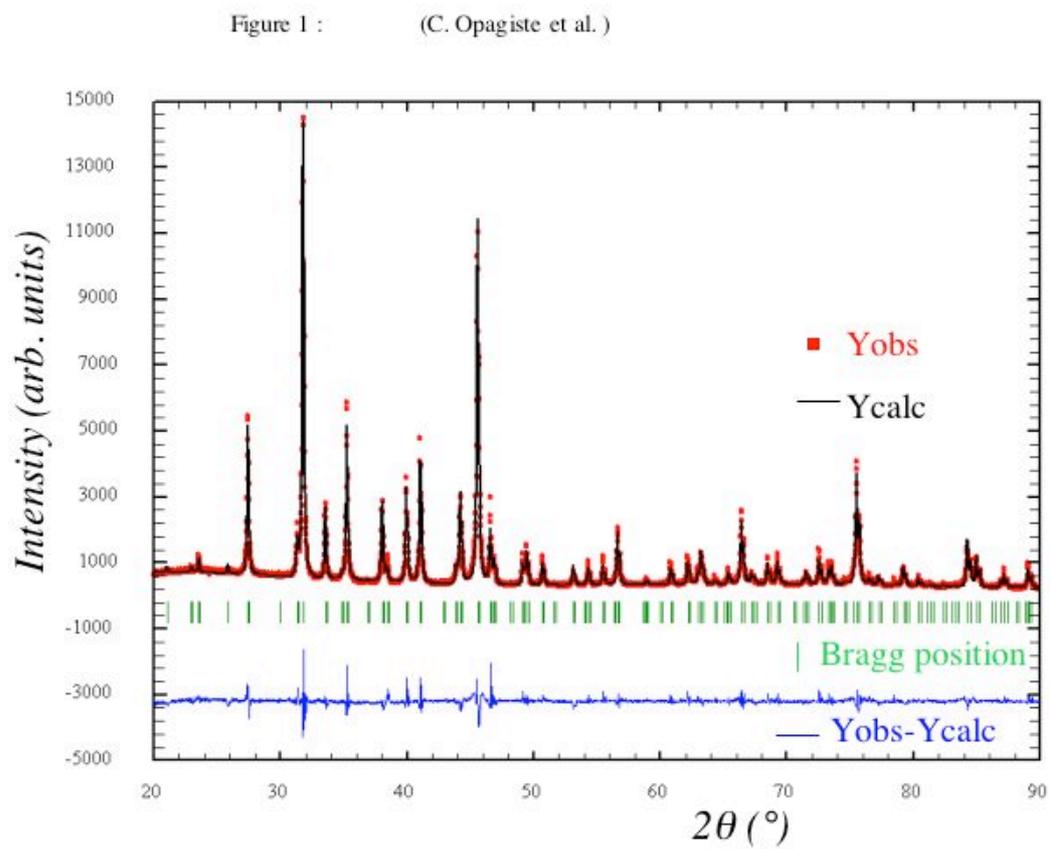

Figure 1 :          (C. Opagiste et al. )

**figure 2**  (C. Opagiste et al.)

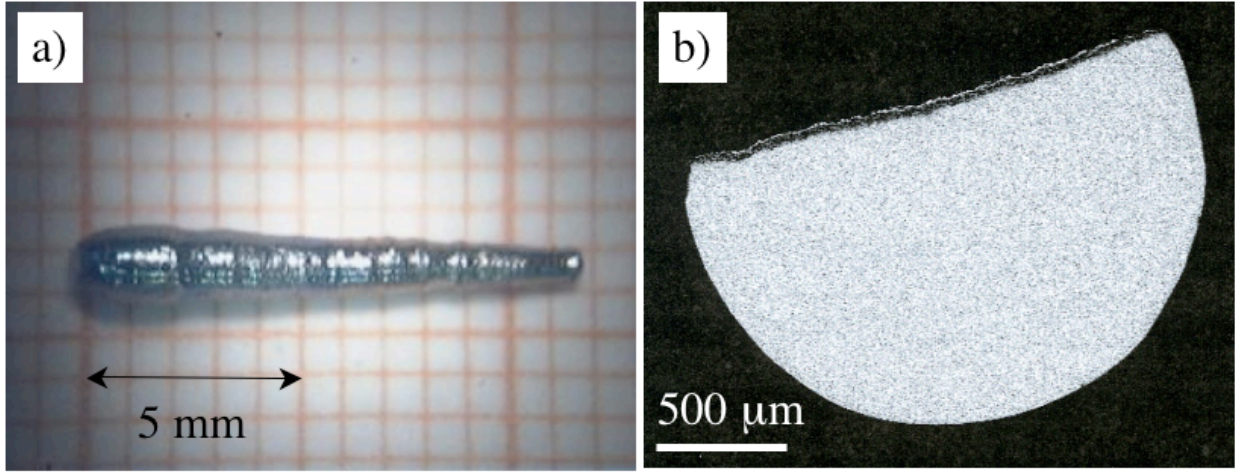

**Figure 3** (C. Opagiste et al.)

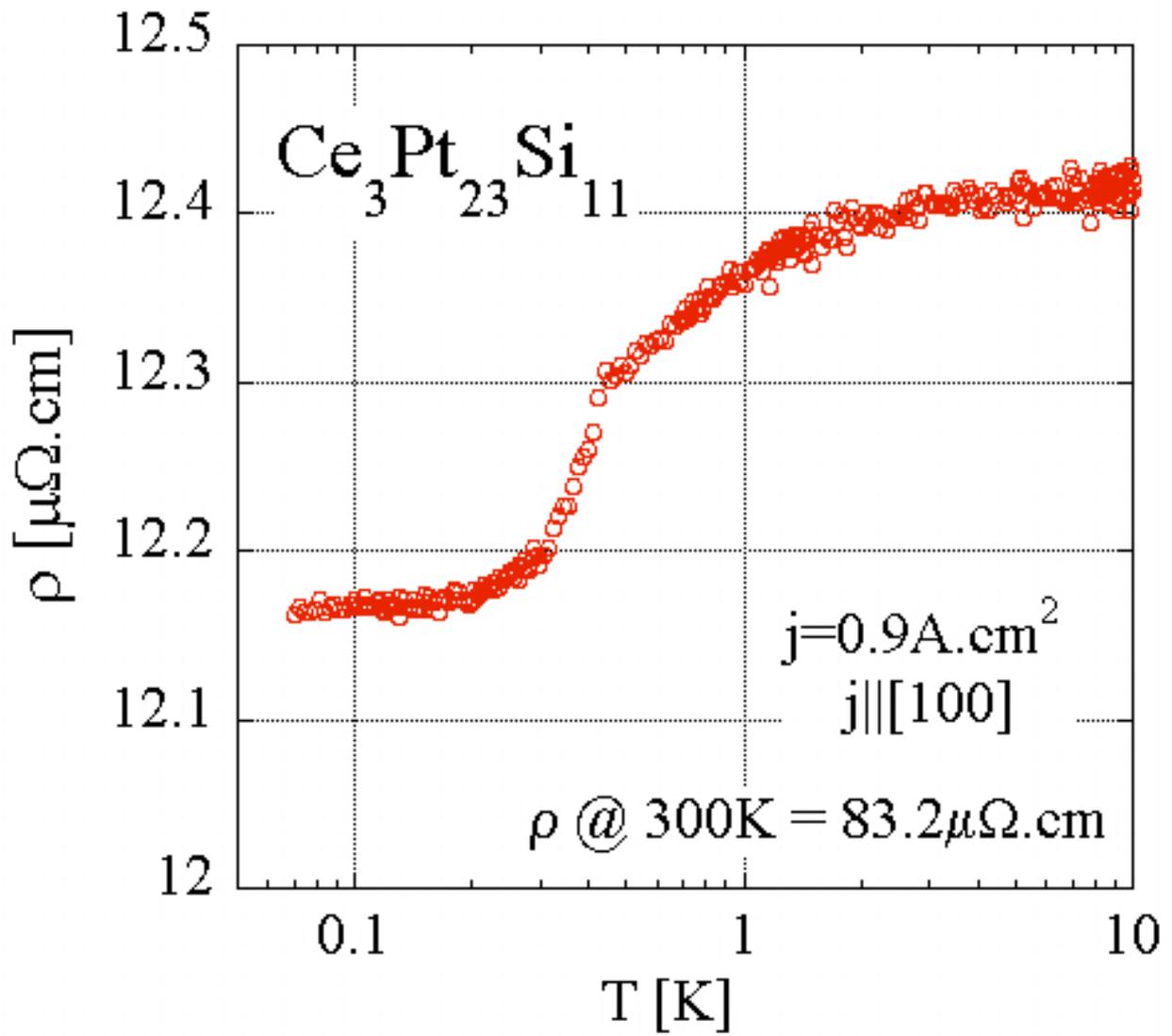

**figure 4** (C. Opagiste et al.)

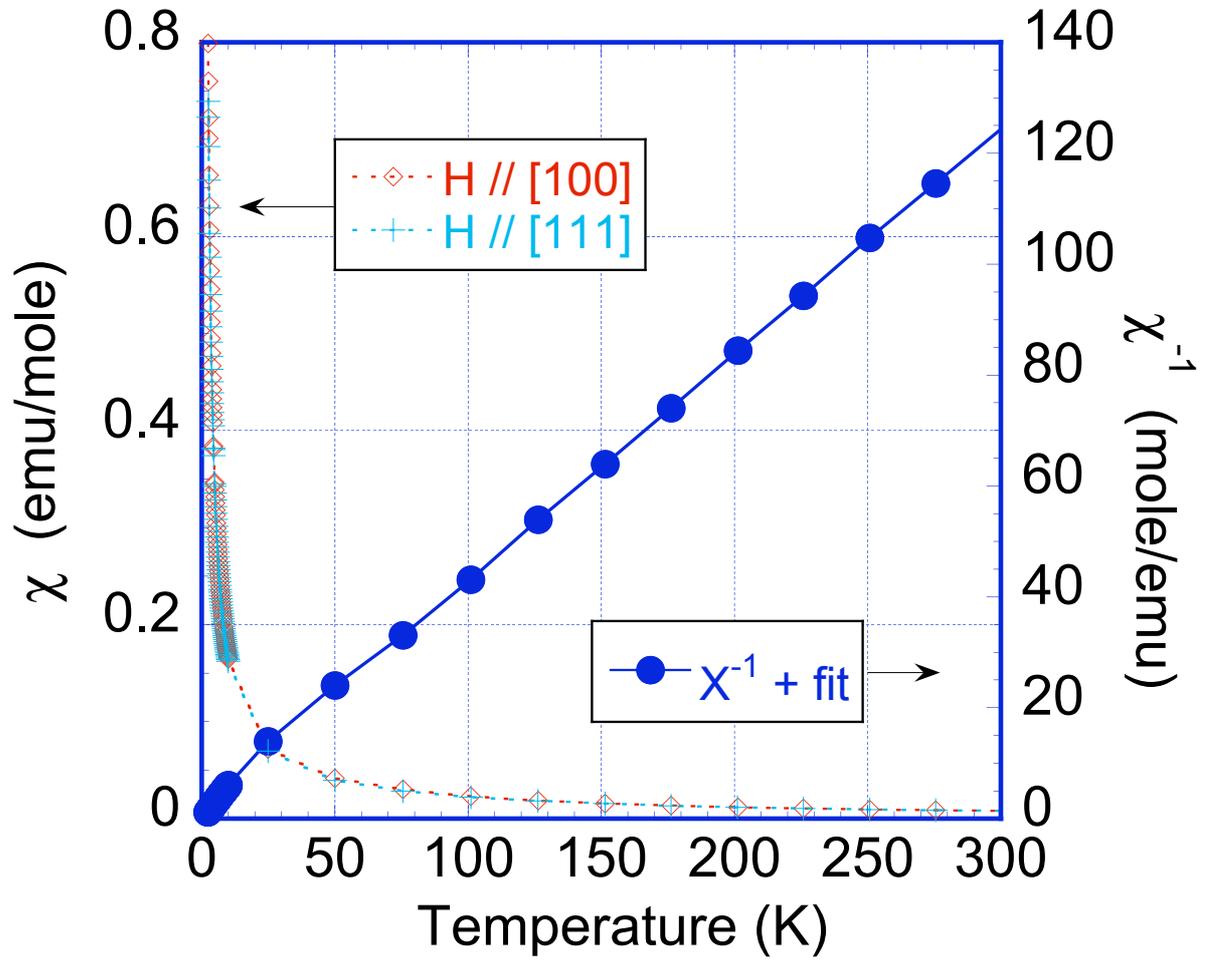

**figure 5** (C. Opagiste et al.)

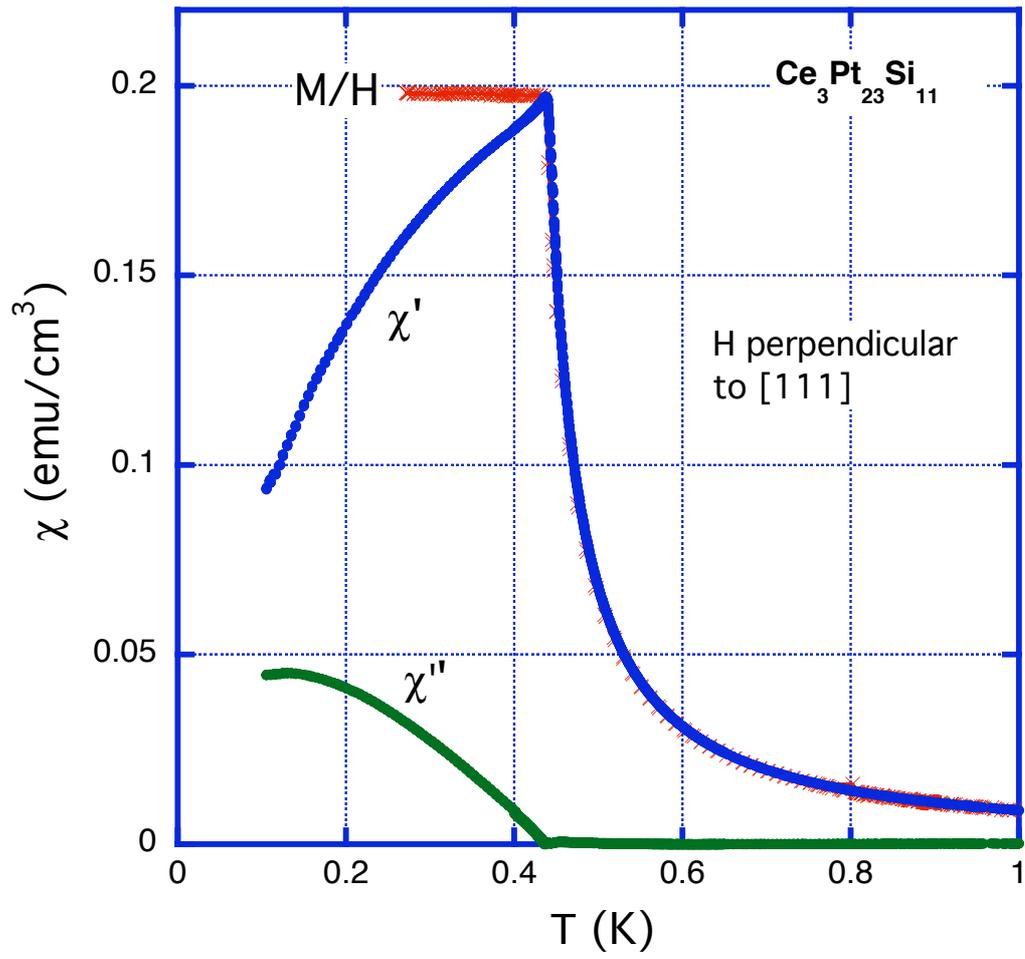

**figure 6** (C. Opagiste et al.)

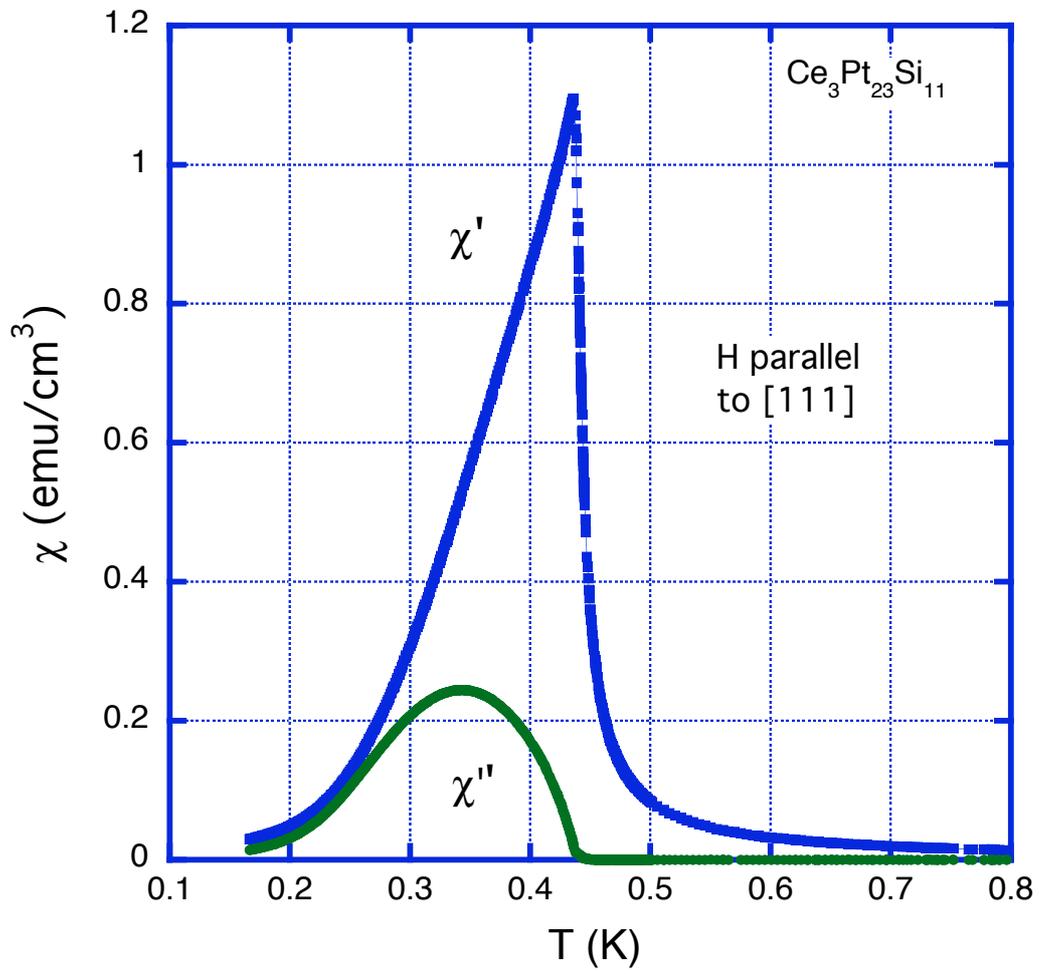

**figure 7**  (C. Opagiste et al.)

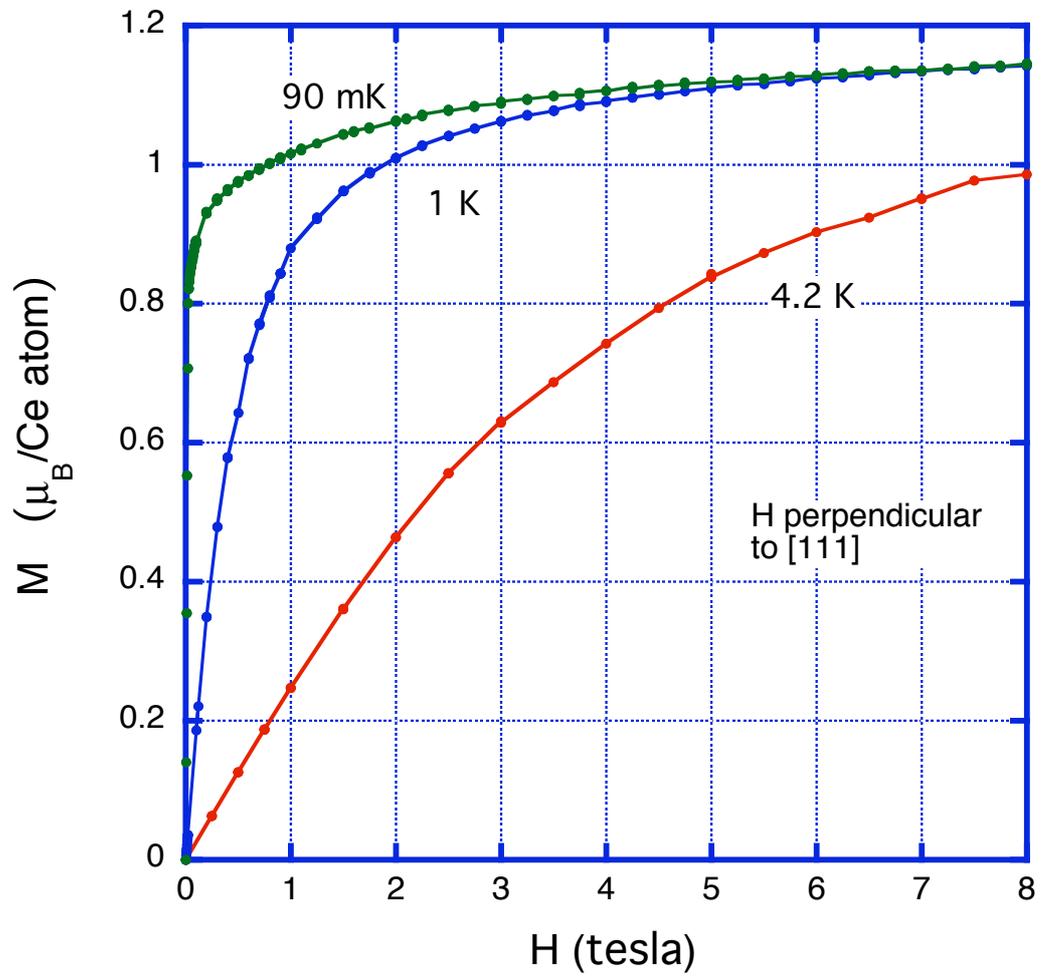

**figure 8** (C. Opagiste et al.)

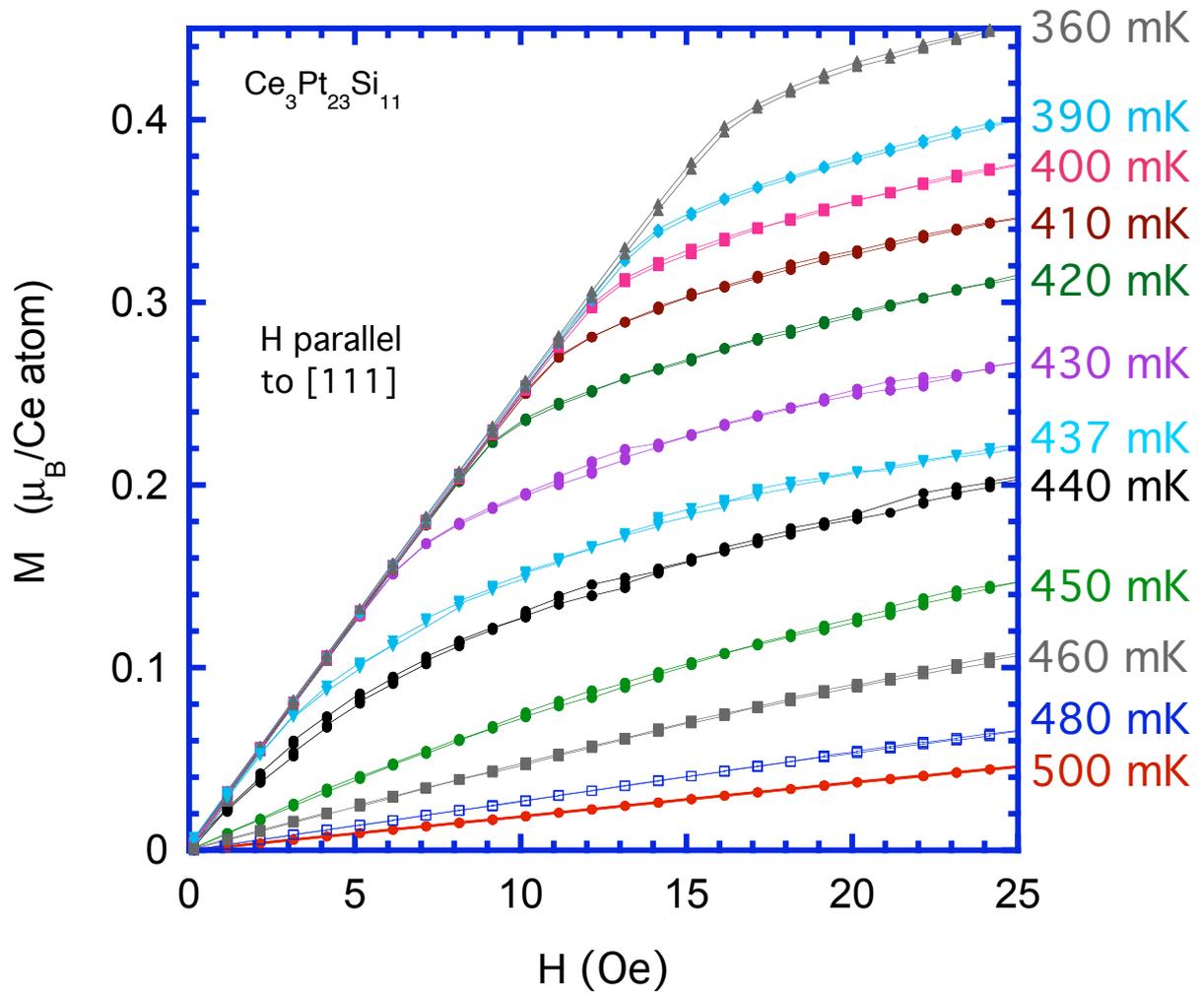

**figure 9** (C. Opagiste et al.)

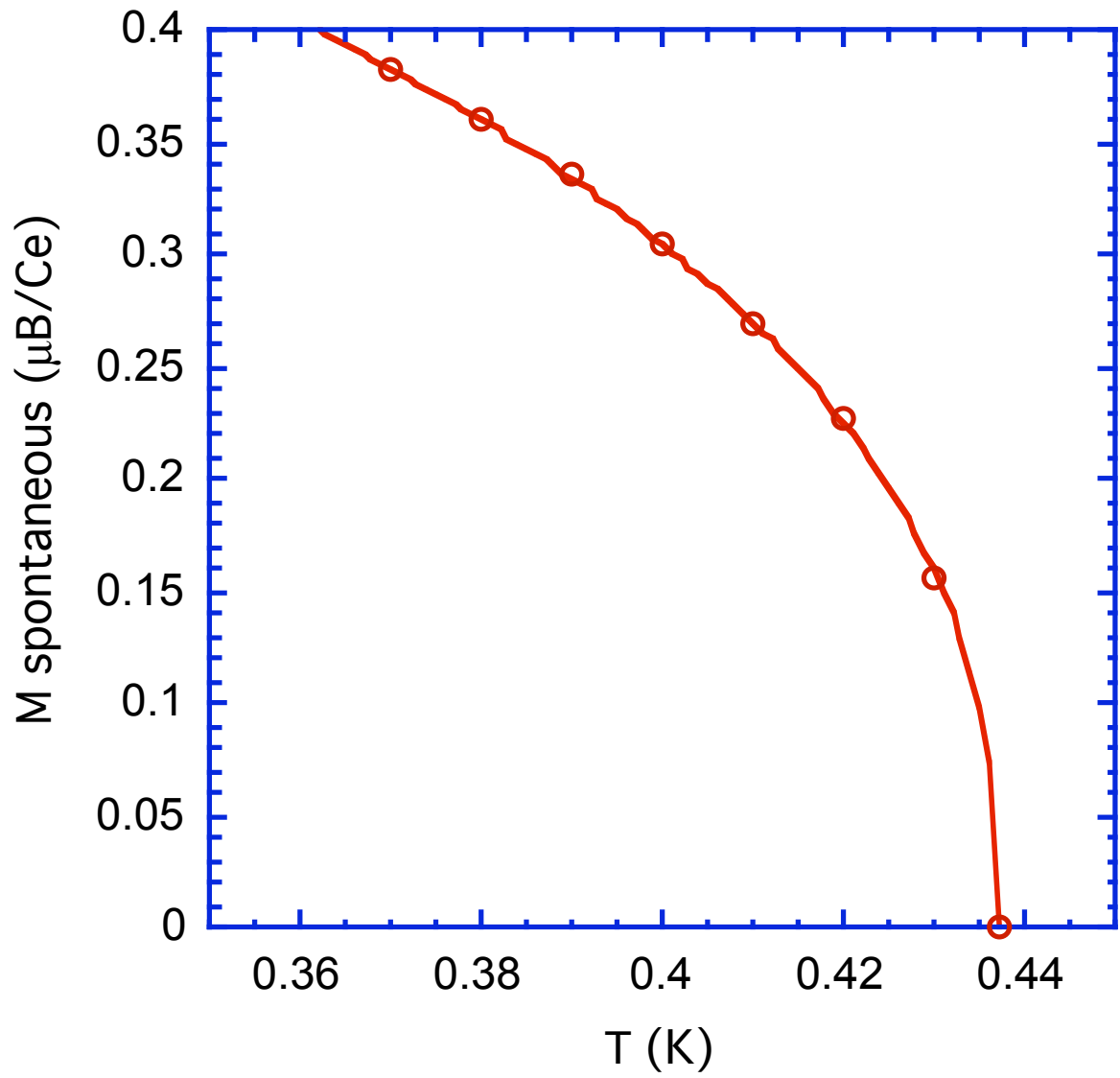

**Table 1** (C. Opagiste et al.)

Quantitative microprobe analysis of our single crystal (the 3$^{rd}$ decimal is given in parentheses).

| Element | Atomic concentration (%) | Standard deviation (At. %) | Atomic concentration (%) for $Ce_3Pt_{23}Si_{11}$ | Atomic concentration (%) for $Ce_2Pt_{15}Si_7$ |
|---|---|---|---|---|
| Ce | 8.16(3) | 0.32(0) | 8.11 | 8,33 |
| Pt | 61.25(1) | 0.09(4) | 62.16 | 62,50 |
| Si | 30.58(6) | 0.16(9) | 29.73 | 29,17 |

## Table 2 (C. Opagiste et al.)

Crystallographic and experimental data and agreement factors for the structure refinement of $Ce_3Pt_{23}Si_{11}$

| | |
|---|---|
| Space group | *Fm-3m* |
| Lattice constant (Å) | 16.901(2) |
| Cell Volume (Å$^3$) | 4838.8 |
| Calculated Density (g.cm$^{-3}$) | 14.316 |
| Z | 8 |
| Wavelength AgKα (Å) | 0.5608 |
| $\mu$ (mm$^{-1}$) | 75.069 |
| maximum θ | 35° |
| Number of measured reflections (observed/all) | 12100 /20392 |
| Unique reflections (observed/all) | 897/1138 |
| Rint (observed/all) | 8.47/8.97 |
| Rwobs/Rwall | 4.15/4.43 |
| g.o.f.obs/ g.o.f.all | 2.40/2.27 |
| number of refined parameters | 30 |
| giso, extinction correction type I (lorentzian distrib.) | 0.00063(5) |

## Table 3 (C. Opagiste et al.)

Atomic coordinates and isotropic displacement parameters for $Ce_3Pt_{23}Si_{11}$

| Atom | Occup. | x | x | x | Uiso (Å$^2$) |
|---|---|---|---|---|---|
| **Pt1** | 1 | 0.08300(2) | 0.08300(2) | 0.08300(2) | 0.00717(7) |
| **Pt2** | 1 | 0.30841(3) | 0.30841(3) | 0.30841(3) | 0.00763(7) |
| **Pt3** | 1 | 0.5 | 0.12573(5) | 0 | 0.00872(14) |
| **Pt4** | 1 | 0.415675(17) | 0.084325(17) | 0.24817(2) | 0.00835(8) |
| **Ce5** | 0.989(7) | 0.25 | 0.25 | 0 | 0.0103(2) |
| **Si6** | 1 | 0.5 | 0.3290(4) | 0 | 0.0072(11) |
| **Si7** | 1 | 0.16529(19) | 0.16529(19) | 0.16529(19) | 0.0088(5) |
| **Si8** | 1 | 0.39384(19) | 0.39384(19) | 0.39384(19) | 0.0083(5) |

**Table 4**  (C. Opagiste et al.)

Harmonic displacement parameters for $Ce_3Pt_{23}Si_{11}$

| Atom | U11 | U22 | U33 | U12 | U13 | U23 |
|---|---|---|---|---|---|---|
| **Pt1** | 0.0072(1) | 0.0072(1) | 0.0072(1) | -0.0001(1) | -0.0001(1) | -0.0001(1) |
| **Pt2** | 0.0076(1) | 0.0076(1) | 0.0076(1) | -0.0003(1) | -0.0003(1) | -0.0003(1) |
| **Pt3** | 0.0092(2) | 0.0077(3) | 0.0092(2) | 0 | 0 | 0 |
| **Pt4** | 0.0082(1) | 0.0082(1) | 0.0086(2) | 0.0012(1) | -0.00022(8) | 0.00022(8) |
| **Ce5** | 0.0106(4) | 0.0106(4) | 0.0097(5) | 0.0001(4) | 0 | 0 |
| **Si6** | 0.007(1) | 0.007(2) | 0.007(1) | 0 | 0 | 0 |
| **Si7** | 0.0088(9) | 0.0088(9) | 0.0088(9) | -0.002(1) | -0.002(1) | -0.002(1) |
| **Si8** | 0.0083(9) | 0.0083(9) | 0.0083(9) | -0.001(1) | -0.001(1) | -0.001(1) |